\documentclass[%
 reprint,
 amsmath,amssymb,
 aps,prl
]{revtex4-1}

\usepackage{graphicx}
\usepackage{dcolumn}
\usepackage{bm}
\usepackage{epsfig}
\usepackage{color}

\newcommand{\ket}[1]{\ensuremath{| #1 \rangle}}
\newcommand{\bra}[1]{\ensuremath{\langle #1 |}}
\newcommand{\ave}[1]{\ensuremath{\langle #1 \rangle}}
\newcommand{\e}{\mathrm{e}}

\begin{document}

\title{Beyond spectral homodyne detection: complete quantum measurement of spectral modes of light}

\author{F. A. S. Barbosa$^1$, A. S. Coelho$^1$, K. N. Cassemiro$^2$, P. Nussenzveig$^1$, C. Fabre$^3$, M. Martinelli$^1$, and  A. S. Villar$^2$}

\affiliation{$^1$Instituto de F\'\i sica, Universidade de S\~ao Paulo, P.O. Box 66318, 
05315-970 S\~ao Paulo, SP, Brazil \\ 
$^2$Departamento de F\'\i{}sica, Universidade Federal de Pernambuco, 50670-901 Recife, PE, Brazil \\
$^3$Laboratoire Kastler Brossel, Case 74, Universit\'e Pierre et Marie Curie, 4 Place Jussieu, 75252 Paris Cedex 05, France}

\begin{abstract}
Spectral homodyne detection, a widely used technique for measuring quantum properties of light beams, cannot retrieve all the information needed to reconstruct the quantum state of spectral field modes. We show that 
full quantum state reconstruction can be achieved with the alternative measurement technique of resonator detection. We experimentally demonstrate this difference by engineering a quantum state with features that go undetected by homodyne detection but are clearly revealed by resonator detection. 
\end{abstract}

\maketitle

Quantum states of light offer novel capabilities for information exchange and processing that are actively investigated. These range from increased security to quantum data compression and quantum teleportation. In order to fully harness such capabilities, it is necessary to coherently exploit the larger configuration space of multimode fields~\cite{leuchsPRL98,pfisterPRL08,lettEntangledimagesScience08,pklamNatPhot09,trepsPRL12,paullettHundredsPRL12,katiPRL12}. Quantum states can then be entangled over many modes and present useful multimode quantum correlations. Spectral modes, with well defined frequencies, are particularly interesting in this respect as they can be easily separated by spectral filtering. 

To determine the spectral content of quantum states of light, homodyne techniques stand as the most important class of quantum measurements available in the continuous variable regime~\cite{yuenshapiroHD80,yuenBalancedOL83,shapiroHD85,yurkeWidebandPRA85,lvovskyraymerRMP09}. By interfering the field of interest with a suitable classical field, the local oscillator (LO), they provide direct access to the phase space distribution of field quadratures.

The detection of spectral modes is usually performed with \textit{spectral homodyne detection}. In this case, a spectrally narrow laser field constitutes the LO, and the detected photocurrent 
undergoes Fourier analysis, its power spectrum providing all the accessible information about the quantum state. The technique has been employed in 
observations of quantum noise squeezing, phase space reconstruction of a squeezed quantum state, and EPR-type entanglement in spectral modes~\cite{slusherFirstSQZPRL85,oupereirakimblePRL92,raymertomoPRL93,mlynekNature97}. 
 More recently, it has allowed the teleportation of quantum correlations as well as partial realization of other quantum information protocols~\cite{furusawaTeleportNature04,pengEntSwapPRL04,furusawaClusterPRL11}. It has been noted however that homodyne detection does not yield a complete measurement of the quantum state of light spanning over the two sidebands modes which contribute to quantum noise~\cite{cavesAmplifiersPRA82,levenson4modesqzPRL87,ralphSinglePhotonSidebandsPRA08}. 

An alternative measurement technique, here called `resonator detection', consists of a `self-homodyne' technique which employs an optical resonator to manipulate the spectral modes prior to detection. This technique allows the measurement of quantum noise in experimental situations in which an external LO is not readily available~\cite{villarPRL05,coelhoScience09}. That feature is thought to be the only advantage of resonator detection, which is usually considered to be equivalent to the spectral homodyne technique as a quantum measurement~\cite{levensonAPB83,levensonOL85,galatolaOptcommun91,villarAJP08}. 

In this Letter, we show that resonator detection indeed provides a complete joint measurement of the two-sideband mode quadrature quantum fluctuations and correlations, allowing us to completely reconstruct the two-mode quantum state. It is therefore more powerful as a quantum measurement than the spectral homodyne detection. We provide an experimental illustration of this property by exhibiting two Gaussian quantum states which are perfectly distinguishable with resonator detection whereas they appear indistinguishable with homodyne detection. 

Photodetection by light absorption is described by the measurement operator $\hat I(t) = \hat E^-(t)\hat E^+(t)$, where the positive and negative frequency parts of the electric field operator form a mode continuum~\cite{mandelwolfBook}, 
\begin{equation}
\hat E^+(t)=\int\! d\omega \,\mathrm{e}^{-i\omega t}\hat a_\omega, \quad \hat E^-(t) = \left(\hat E^+(t)\right)^\dag.
\end{equation}
Field amplitude $\hat p_\omega$ and phase $\hat q_\omega$ quadrature observables, fulfilling the canonical commutation relation $[\hat p_\omega,\hat q_{\omega'}]=2i\delta(\omega-\omega')$, satisfy $\hat a_\omega = (\hat p_\omega + i \hat q_\omega)/2$.

Homodyne detection employs as input field a LO mode, i.e. a coherent state $\ket{\alpha}_{\omega_0}$ (at frequency $\omega_0$), to amplify the quantum noise stemming from modes in its frequency vicinity and possessing the quantum state $\hat\rho$ of interest. In the balanced detection configuration, the associated quantum measurement is represented by the operator  
\begin{equation}
\delta \hat I(t) = \alpha^*\hat a(t)+ \alpha\hat a^\dag(t),  
\end{equation}
where $\hat a(t)=\int d\omega\,\e^{-i(\omega-\omega_0) t}\,\hat a_{\omega}$, and the frequency integral is limited around $\omega_0$ by detection bandwidth. 

Spectral resolution of field modes is achieved by Fourier analysis of $\delta\hat I(t)$. Experimentally, it is performed by mixing the photocurrent with a sinusoidal electronic reference at frequency $\Omega$ and integrating the result for a time compatible with the desired spectral resolution. The following operator represents the quantum measurement of the spectral photocurrent fluctuations:
\begin{equation}
\label{eq:spectralphotocu}
\delta \hat I_\Omega=\alpha^*\,\hat a_{\omega_0+\Omega}+\alpha\,\hat a^\dag_{\omega_0-\Omega}= |\alpha|(\hat I_{\cos{}} + i\hat I_{\sin{}}),
\end{equation}
$ \hat I_{\cos{}}$ and $i\hat I_{\sin{}}$ being the two Hermitian operators associated with the cosine and sine photocurrent observables~\cite{yuenshapiroHD80,cavesAmplifiersPRA82}. They are given by 
\begin{align}
\label{eq:HDIcosIsin}
\hat I_{\cos{}}(\varphi) &= \cos\varphi\;\hat p_++\sin\varphi\;\hat q_+\equiv \hat X_+(\varphi),\\
\hat I_{\sin{}}(\varphi) &= -\sin\varphi\;\hat p_- + \cos\varphi\;\hat q_- \equiv \hat X_-(\varphi+\pi/2),\nonumber
\end{align}
where the LO phase $\varphi$ (in $\alpha = |\alpha|\mathrm{e}^{i\varphi}$) determines the direction of quadrature observation in phase space. Modes labeled by subscripts $+$ and $-$ are the symmetric and anti-symmetric coherent combinations of spectral sideband modes $\omega_0\pm\Omega$, in the form
\begin{equation}
\label{eq:ppmqpm}
\hat p_\pm ={\textstyle\frac{1}{\sqrt{2}}} (\hat p_{\omega_0+\Omega} \pm \hat p_{\omega_0-\Omega}) ,\; \hat q_\pm ={\textstyle\frac{1}{\sqrt{2}}} (\hat q_{\omega_0+\Omega} \pm \hat q_{\omega_0-\Omega}).
\end{equation}
They represent the quadrature operators of modes naturally associated with spectral homodyne detection.

The measurement operators of Eq.~(\ref{eq:HDIcosIsin}) are each single-mode quadrature observables~\cite{gardinerJMO87}. The cosine photocurrent component refers to the symmetric mode and the sine component to the anti-symmetric mode. The LO phase controls the direction of observation in phase space for \textit{both} individual modes, which are thus connected to one another: It is not possible to rotate phase spaces independently. Hence only two-mode correlations of the type $\ave{\hat X_+(\varphi)\hat X_-(\varphi+\pi/2)}$ can be accessed, while moments of the form $\ave{\hat X_+(\varphi)\hat X_-(\varphi)}$ are missing. 

Consider $\mathcal{E}_{\omega_0\pm\Omega}=\frac{1}{2}(\Delta^2\hat p_{\omega_0\pm\Omega}+\Delta^2\hat q_{\omega_0\pm\Omega})-1$ as the energy present in the quantum fluctuations of each spectral mode. The missing two-mode moment then reads as $\mathcal{E}_{\omega_0+\Omega}-\mathcal{E}_{\omega_0-\Omega}$. In the spectral modal basis, it has a clear meaning: energy imbalance. With only one controllable parameter, homodyne detection is blind to two-mode energy asymmetry, since it treats spectral modes in a perfectly indistinguishable manner according to Eq.~(\ref{eq:HDIcosIsin}).

In most experiments, the phases of the optical field and the electronic oscillator are not locked to each other. Thus, only the photocurrent spectral noise power $S_\mathrm{HD} = |\alpha|^2(\Delta^2\hat I_{\cos{}}+\Delta^2\hat I_{\sin{}})/2$ carries quantum state information~\cite{barbosaPRA2be}. Its expression in terms of symmetric and anti-symmetric modes reads as
\begin{align}
\label{eq:imixhd}
&\qquad S_\mathrm{HD}(\varphi) 
={\textstyle\frac{1}{2}}\cos^2\!\varphi\left(\Delta^2\hat p_++\Delta^2\hat q_-\right)+\\
&+{\textstyle\frac{1}{2}}\sin^2\!\varphi\left(\Delta^2\hat p_-+\Delta^2\hat q_+\right)+{\textstyle\frac{1}{2}}\sin\!2\varphi\left(C_{\hat p_+\hat q_+}-C_{\hat p_-\hat q_-}\right),\nonumber
\end{align}
where $C_{\hat o\hat o'}$
denotes the symmetrized correlation function of quadratures $\hat o$ and $\hat o'$. 
By scanning the LO phase $\varphi$, one can therefore measure only the three  noise dependent coefficients figuring in Eq. (\ref{eq:imixhd}).

Let us now turn to resonator detection. The spectral mode annihilation operator is modified after reflection on the cavity according to 
\begin{equation}
\label{eq:resdetecmodif}
\hat a_{\omega}\longrightarrow r(\Delta_\omega)\,\hat a_{\omega} + t(\Delta_\omega)\,\hat b_{\omega},
\end{equation}
where the resonator reflection $r(\Delta_\omega)$ and transmission $t(\Delta_\omega)=\sqrt{1-r^2(\Delta_\omega)}$ coefficients, which  combine the input quantum field of interest $\hat a_{\omega}$ with a field mode $\hat b_{\omega}$ in vacuum state, are functions of cavity detuning $\Delta_\omega=(\omega-\omega_c)/\gamma$ relative to the cavity bandwidth $\gamma$~\cite{villarAJP08}. In addition, the resonator performs the transformation $\alpha\longrightarrow \alpha_r = r(\Delta)\,\alpha$ on the mean value of the input field of frequency $\omega_0$ and amplitude $\alpha$. The detuning $\Delta = (\omega_0-\omega_c)/\gamma$ is a parameter which is experimentally controllable by scanning the cavity length. The intensity of the reflected field is then measured and Fourier analyzed, yielding $\Delta$ dependent photocurrent spectral components $\hat J_\mathrm{cos}$ and $\hat J_\mathrm{sin}$ given by
\begin{align}
\label{eq:Jcossin}
\hat J_\mathrm{cos} =&\;x_{\Omega}\,\hat p_{\Omega}+y_{\Omega}\,\hat q_{\Omega}+x_{-\Omega}\,\hat p_{-\Omega} +y_{-\Omega}\,\hat q_{-\Omega} +\hat J_u, \\
\hat J_\mathrm{sin} =&-y_{\Omega}\,\hat p_{\Omega}+x_{\Omega}\,\hat q_{\Omega}+y_{-\Omega}\,\hat p_{-\Omega} -x_{-\Omega}\,\hat q_{-\Omega}+\hat J_v,\nonumber
\end{align}
where mode notation has been compacted, $\omega_0\pm\Omega\rightarrow \pm\Omega$, and coefficients $x_{\pm\Omega}$ and $y_{\pm\Omega}$ are real functions of $\Delta$, 
\begin{equation}
x_{\pm\Omega}+iy_{\pm\Omega}=R_\Omega=\frac{r(\Delta)}{|r^*(\Delta)|}\;r(\Delta\pm\Omega/\gamma)
\end{equation}
Operators $\hat J_u$ and $\hat J_v$ stand for vacuum contributions related to cavity losses. We see in expression (\ref{eq:Jcossin}) that, unlike homodyne detection, the resonator technique inherently gives more information about the two spectral modes, from which it is possible to derive the whole two-mode covariance matrix~\cite{ralphSeparatingSidebandsJOB02,barbosaPRA2be}, and therefore perform the reconstruction of any two-mode Gaussian state.

The corresponding spectral noise power is 
\begin{align}
\label{eq:SRDenergy}
S_\mathrm{RD}(\Delta) &=  g_+(\Delta)(\mathcal{E}_\Omega+\mathcal{E}_{-\Omega}+4)+g_-(\Delta)(\mathcal{E}_\Omega-\mathcal{E}_{-\Omega})+\nonumber\\
& +g_r(\Delta) (\Delta^2\hat p_++\Delta^2\hat q_--\Delta^2\hat p_--\Delta^2\hat q_+) + \nonumber\\
& +g_i(\Delta)\left(C_{\hat p_+\hat q_+}-C_{\hat p_-\hat q_-}\right).
\end{align}
where $g_\pm=|R_\Omega|^2\pm|R_{-\Omega}|^2$ stand for the sum or subtraction of resonator transmission spectra, and $g_r$ and $g_i$ are real functions of $\Delta$ defined by $g_r+ig_i=R_\Omega R_{-\Omega}/2$. 

The different $g$ being known functions of $\Delta$, it is possible to retrieve their four coefficients from a least squares fit to the experimental data.  
Resonator detection offers in particular the ability to pinpoint quantum states showing strong energy imbalance between spectral sideband modes. The reason for that is the introduction not only of dispersion by the optical resonator, but also of frequency-dependent modal attenuation.

\begin{figure}[ht]
\includegraphics[width=0.95\linewidth]{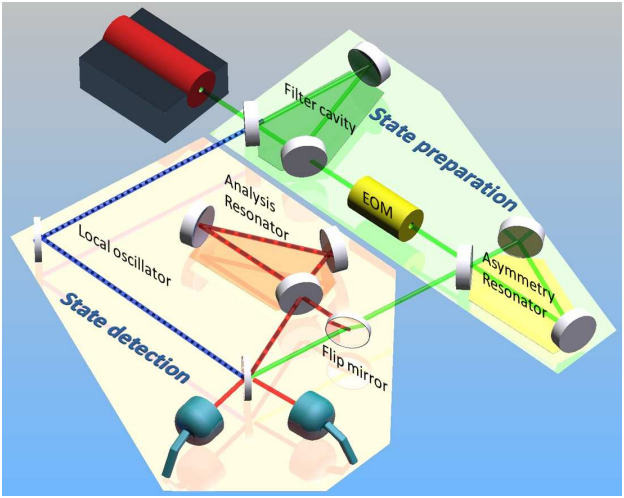}
\caption{(Color online) Experimental setup. State preparation (solid green beam): EOM displaces spectral modes and asymmetry resonator replaces one of them by vacuum. State detection: Either homodyne (dotted blue beam) or resonator detection (dashed red beam) are employed (flip mirror). }
\label{fig:setup}
\end{figure}

We generate a quantum state possessing energy imbalance that is likely to benchmark the differences between homodyne and resonator detection.  
In our experiment, such quantum state is produced in two steps, as depicted in Figure~\ref{fig:setup}. Firstly, two spectral modes are displaced with the aid of an electro-optic modulator (EOM) to produce coherent states with complex conjugate amplitudes, $\ket{\psi_1} = \hat D_{\omega_0-\Omega}(\beta^*) \hat D_{\omega_0+\Omega}(\beta) \ket{0} = \ket{\beta^*}_{\omega_0-\Omega} \ket{\beta}_{\omega_0+\Omega}$, where $\beta=|\beta|\mathrm{e}^{i\theta}$. This state vector shows balanced spectral energy distribution. Secondly, an auxiliary optical resonator, called asymmetry resonator, produces energy imbalance between spectral modes by attenuating one of them in reflection. In our case, the asymmetry resonator is locked to resonance with mode $\omega_0-\Omega$, and the quantum state becomes $\ket{\psi_2} = \ket{\beta'{}^*}_{\omega_0-\Omega} \ket{\beta}_{\omega_0+\Omega}$, where $\beta' = \sqrt{R_0'}\,\beta$. 

Quantum state preparation and measurement are illustrated in Fig.~\ref{fig:setup}. The input laser beam at 532~nm is mode cleaned by a filter cavity. The same signal used for EOM modulation is employed as electronic reference in the measurement of quantum noise, with frequency $\Omega = 17$~MHz~\cite{mlynekNature97,darianoOL98}. Spectral analysis of the photocurrent is performed with 600~kHz bandwidth. Each intermediate step $\ket{\psi_1}$ and $\ket{\psi_2}$ for the quantum state preparation of $\hat\rho$ is experimentally verified. Although we employ resonator detection at this stage, homodyne detection finds the same measurement results, since only first-order moments appear and all second-order moments are either shot noise or null. 

\begin{figure}[ht]
\includegraphics[width=0.49\linewidth]{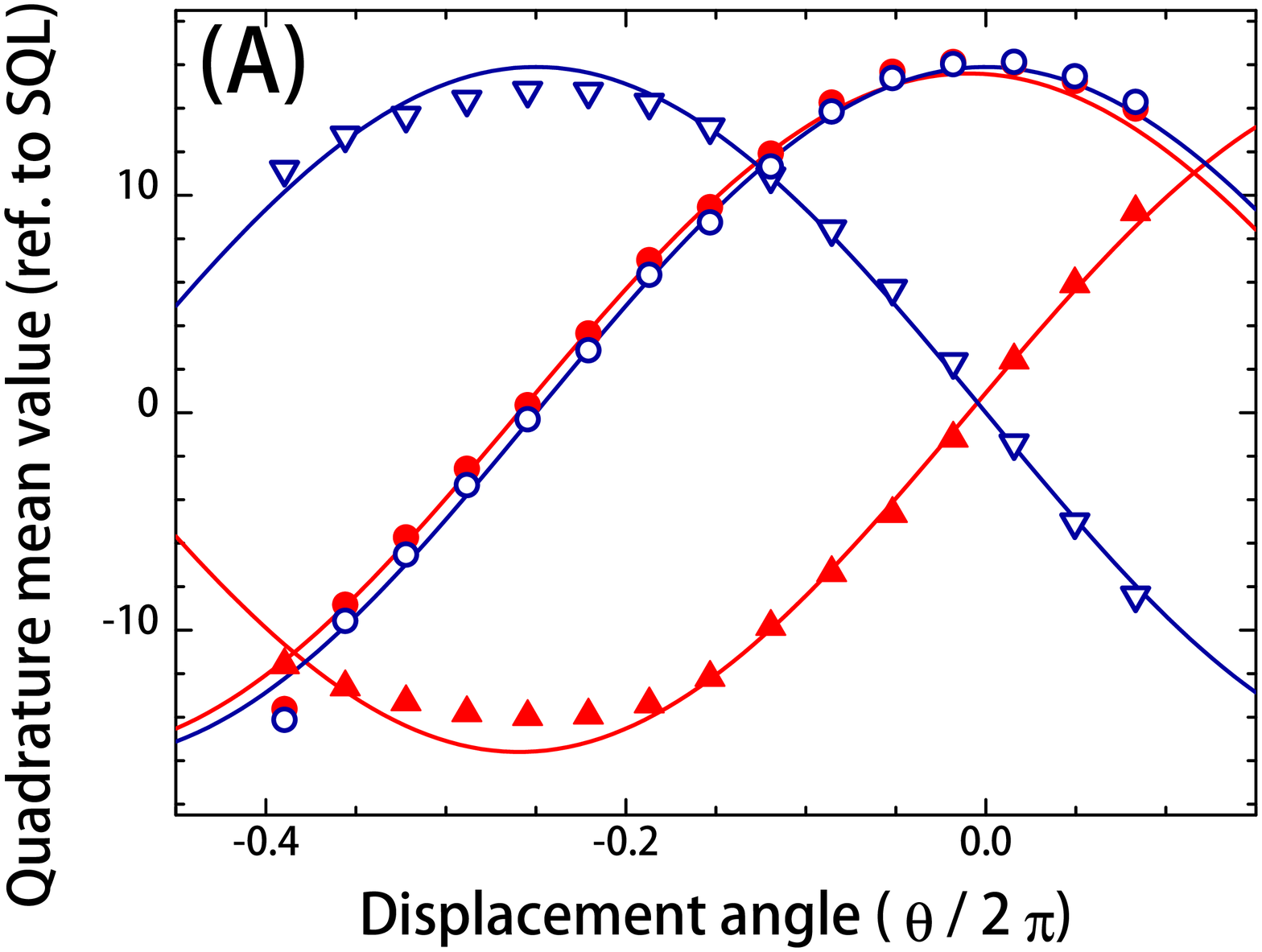}
\includegraphics[width=0.49\linewidth]{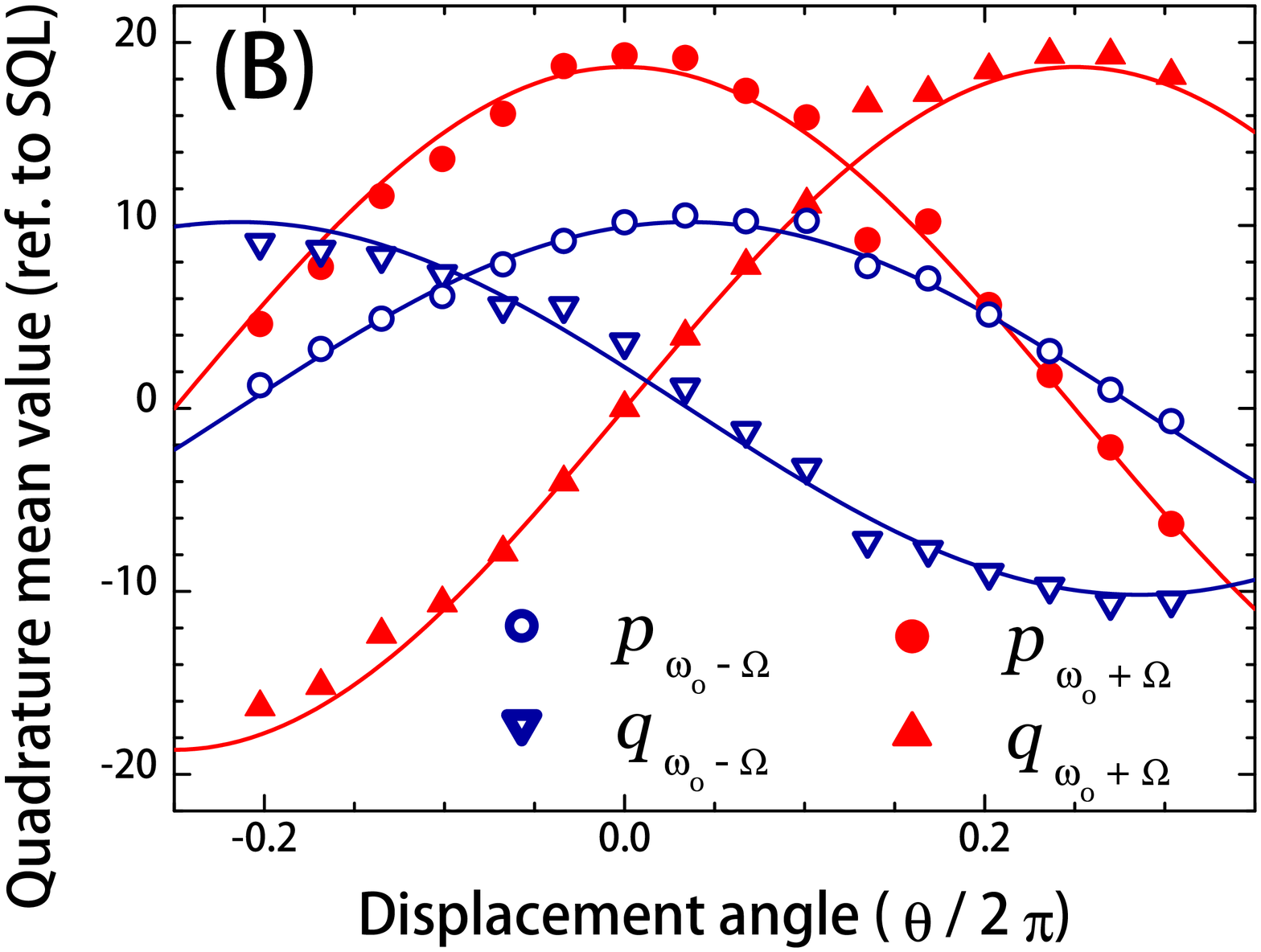}
\\\vspace{5pt}
\includegraphics[width=0.3\linewidth]{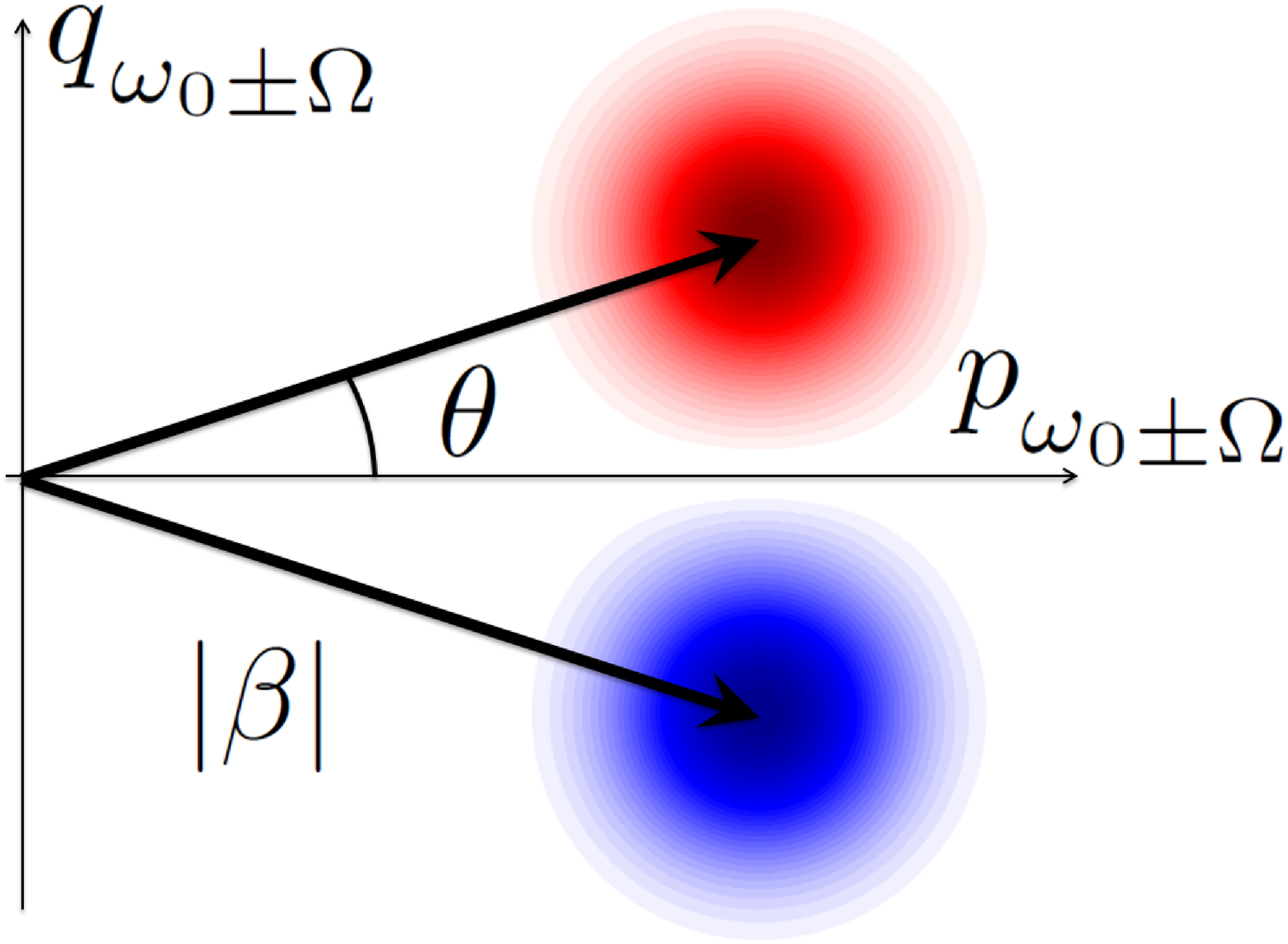}
\hspace{50pt}
\includegraphics[width=0.3\linewidth]{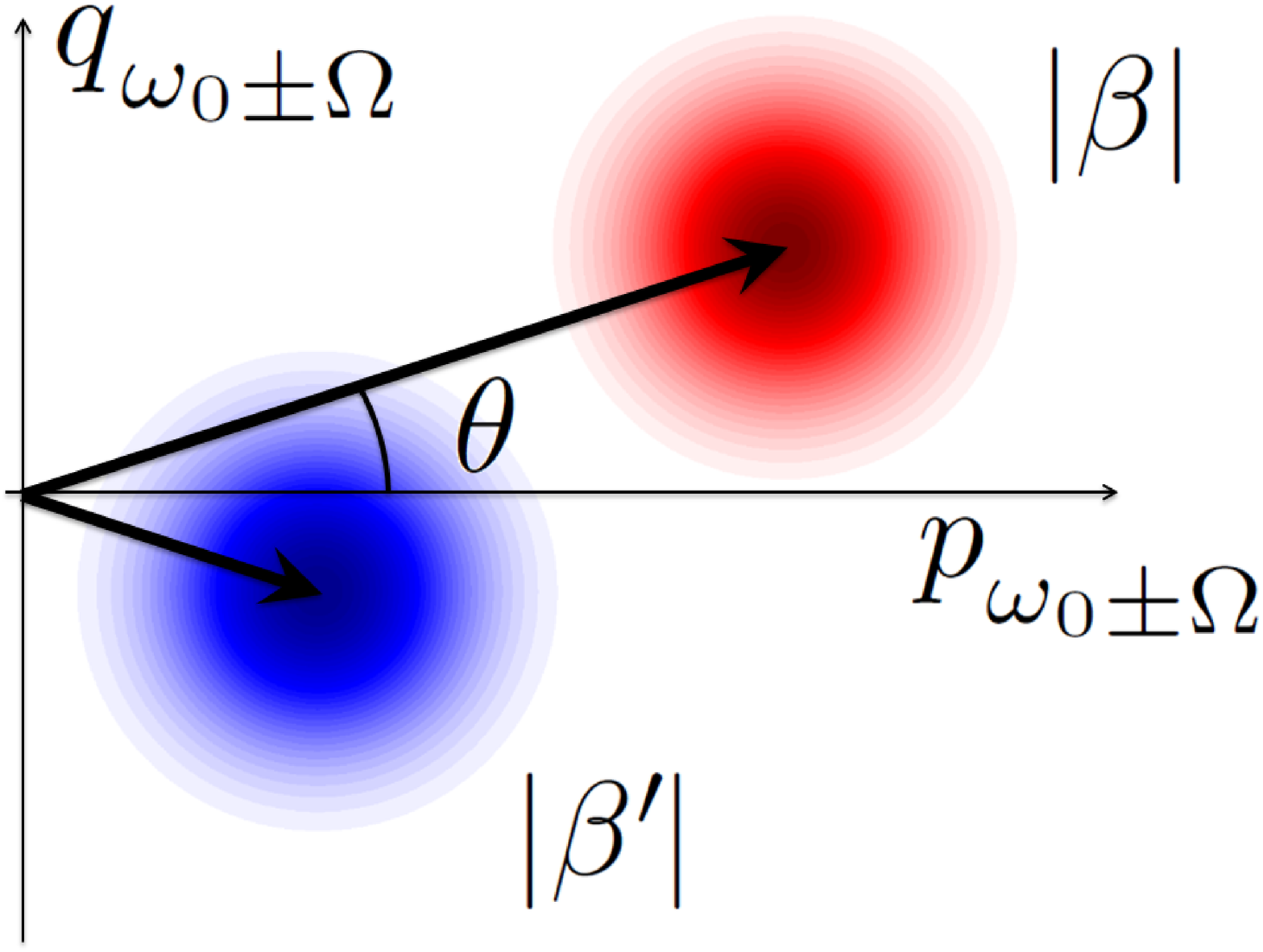}
\caption{(Color online) Expected values of spectral quadrature observables as functions of displacement angle $\theta$. Left: Quantum state $\ket{\psi_1}$ with balanced energy distribution. Right: Quantum state $\ket{\psi_2}$ with spectral energy imbalance. }
\label{fig:data2}
\end{figure}

Data is presented in Fig.~\ref{fig:data2}. In the first step, quadrature state averages on $\ket{\psi_1}$ relative to the standard quantum level (SQL) are shown in Fig.~\ref{fig:data2}A as functions of $\theta$. Quadrature mean values are obtained by theoretical fits of quantum state averages of Eq.~(\ref{eq:Jcossin}) to the data. Amplitude and phase quadratures alternate with $\theta$ in pairs as projections of $\beta=\ave{\hat p_{\omega_0+\Omega}}_1+i\ave{\hat q_{\omega_0+\Omega}}_1$ and $\beta^*=\ave{\hat p_{\omega_0-\Omega}}_1-i\ave{\hat q_{\omega_0-\Omega}}_1$. Complex conjugate displacements are attested by data. 
In the second step, the asymmetry resonator produces state $\ket{\psi_2}$ with energy imbalance. Reflection at resonance is $R_0'=0.12(2)$\%, and bandwidth is 4.1(2)~MHz. Displacement amplitudes for both spectral modes can be observed in Fig.~\ref{fig:data2}. While one spectral mode presents $\ave{\hat p_{\omega_0+\Omega}}_2\approx19$ at maximum, the other shows $\ave{\hat p_{\omega_0-\Omega}}_2\approx10$, attesting energy imbalance ratio of $\mathcal{E}_{\omega_0-\Omega}/\mathcal{E}_{\omega_0+\Omega}\approx 0.28$. 

We seek to create a state with an imbalance in the second-order moments. This is achieved by randomizing the amplitudes and phases of the displacements, by driving the EOM with Gaussian noise as input. This procedure scrambles the displacement mean values and produces field fluctuations with the desired properties. The desired benchmark quantum state 
$\hat\rho = \int d^2\beta \;\mathrm{e}^{-\frac{|\beta|^2}{|\beta_0|^2}}\ket{\psi_2}\bra{\psi_2}$ bearing modal energy imbalance, where $|\beta_0|$ is the typical modulation energy, is thus generated. 
For each single-mode, $\hat\rho$ is a thermal state. 
We then measure the benchmark quantum state $\hat\rho$ with homodyne and resonator detection schemes (Fig~\ref{fig:setup}). For comparison purposes, we also prepare a reference thermal state $\hat\rho_r$ showing balanced energy distribution, by amplitude randomization of $\ket{\psi_1}$ in the form $\hat\rho_r = \int d^2\beta \;\mathrm{e}^{-\frac{|\beta|^2}{|\beta_0|^2}}\ket{\psi_1}\bra{\psi_1}$. 

\begin{figure}[ht]
\raisebox{1.0cm}{\includegraphics[width=0.35\linewidth]{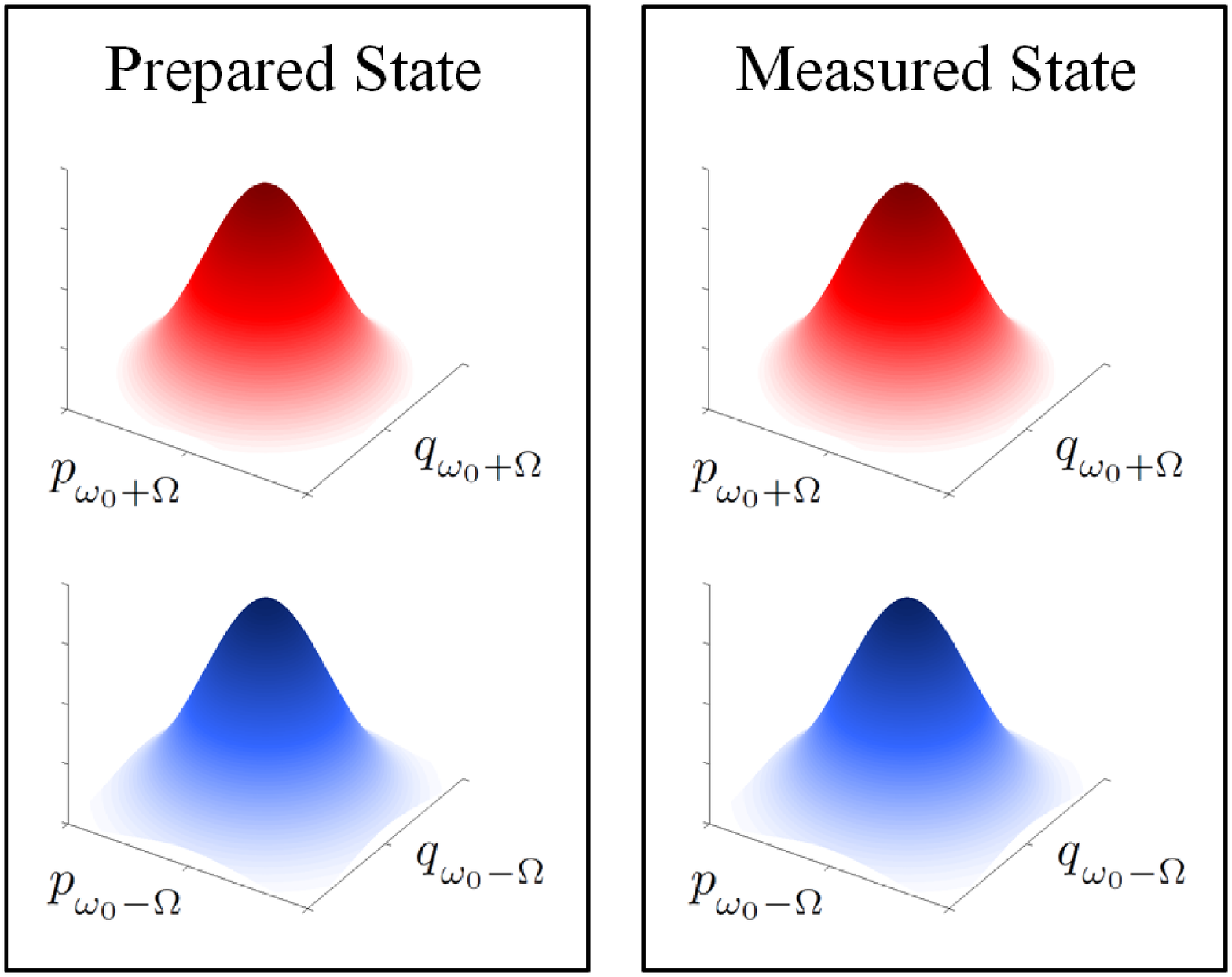}}
\hspace{5pt}
\includegraphics[width=0.6\linewidth]{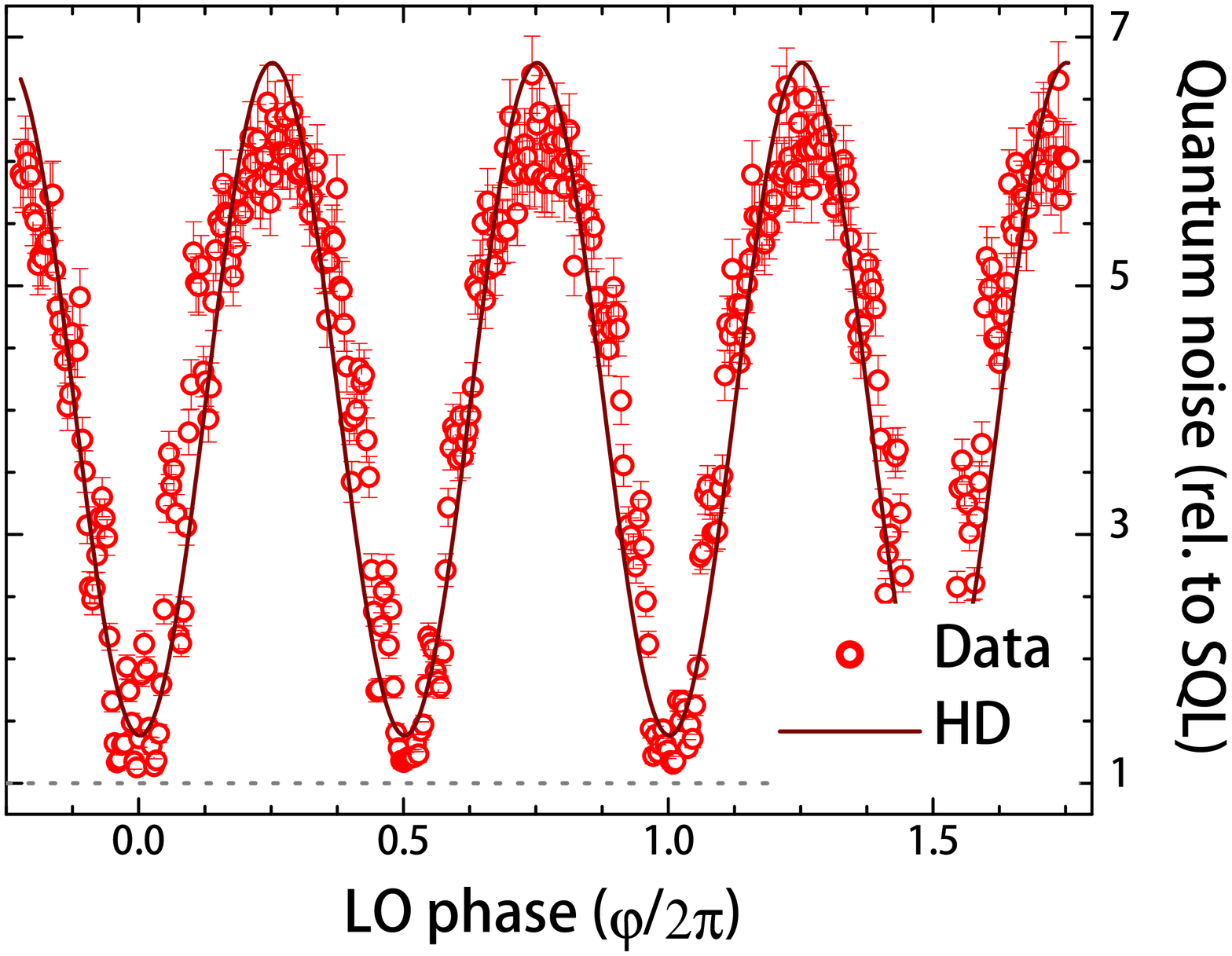}
\newline
\includegraphics[width=0.6\linewidth]{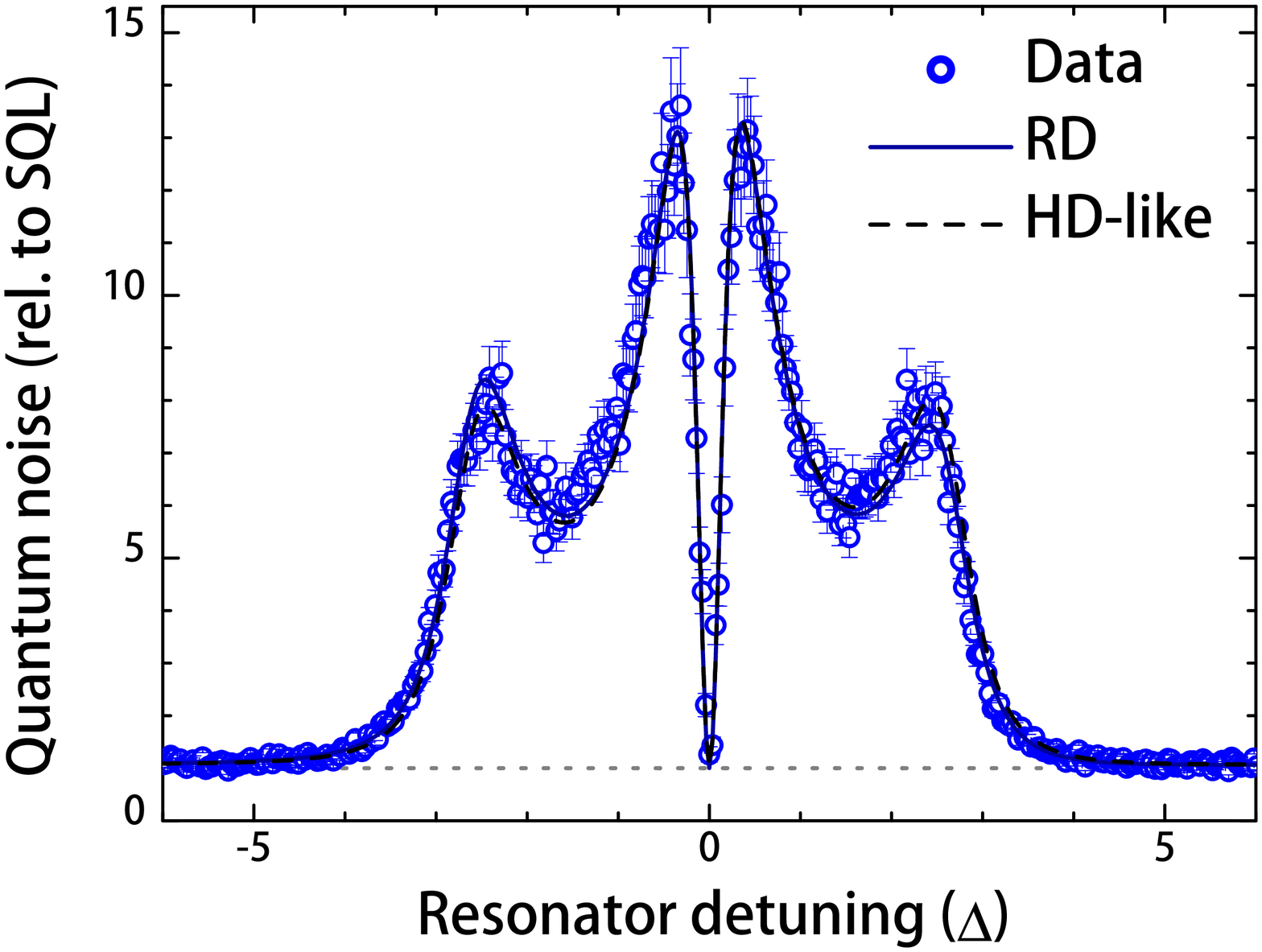}
\hspace{5pt}
\raisebox{1.0cm}{\includegraphics[width=0.35\linewidth]{wigner_insetSym2.eps}}
\caption{(Color online) Measurements of spectral quantum noise produced by the balanced quantum state $\rho_r$. Top: Homodyne detection, as a function of $\varphi$. Bottom: Resonator detection, as a function of $\Delta$. Solid lines (`HD' on top and `RD' on bottom) are theoretical fits to the data. Dashed line on the bottom (`HD-like') represents the theoretical noise curve of Eq.~(\ref{eq:SRDenergy}) as it would appear in resonator detection using only moments obtainable with homodyne detection.}
\label{fig:data1Sym}
\end{figure}

Experimental results for the symmetric state $\hat\rho_r$ using both techniques are presented in Fig.~\ref{fig:data1Sym}. Data with homodyne detection appears at the top right. Fringe contrast exceeds 91\%. The solid line (labeled HD) depicts the theoretical fit of Eq.~(\ref{eq:imixhd}) to the data. Insets depict the Wigner functions of individual spectral modes associated with measured operator moments, yielding a visual representation of the measured single-mode covariance matrices. 

In the bottom left, we observe resonator detection data obtained from a symmetric quantum state produced in a similar manner. The only difference is the increase in mean energy to improve signal to noise ratio. Resonator bandwidth is $\gamma = 6.0(3)$~MHz, mode coupling efficiency is $93.5(1.0)\%$ and $R_0=0.04(2)$. The solid line (labeled RD) represents a theoretical fit of Eq.~(\ref{eq:SRDenergy}) to the data (taking into account that mode coupling efficiency slightly modifies the equation). The dashed line (HD) represents noise as it would be observed with resonator detection in case only moments available to homodyne detection had been retrieved. Both curves provide equally good fits. 

We see in this first series of measurements that both techniques are able to recover correctly the quantum state prepared, as established by the Wigner functions in insets. Resonator and homodyne detection find compatible results for $\hat\rho_r$, meaning that no spectral energy imbalance is necessary to describe the quantum state. 

\begin{figure}[ht]
\raisebox{1.0cm}{\includegraphics[width=0.35\linewidth]{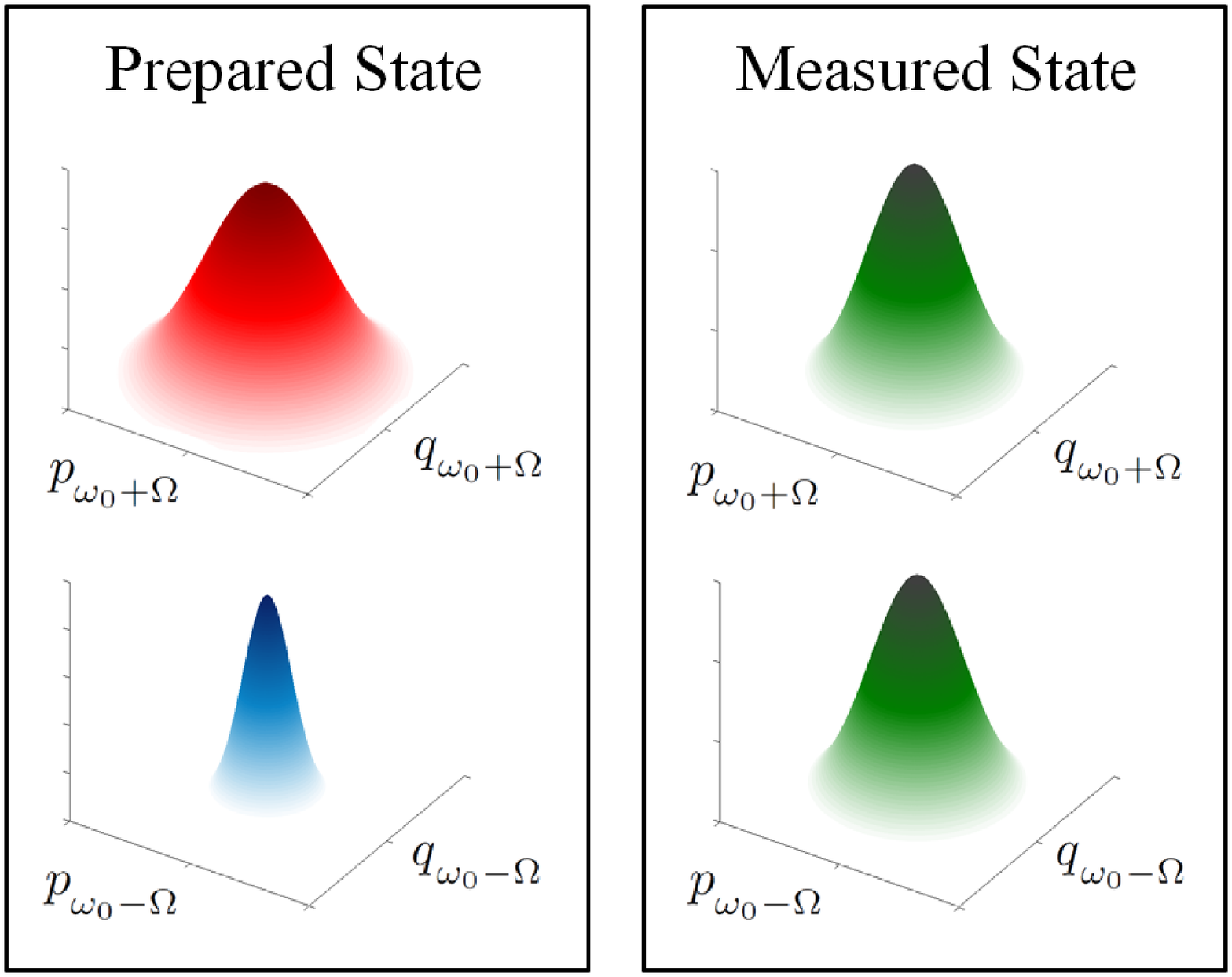}}
\hspace{3pt}
\includegraphics[width=0.6\linewidth]{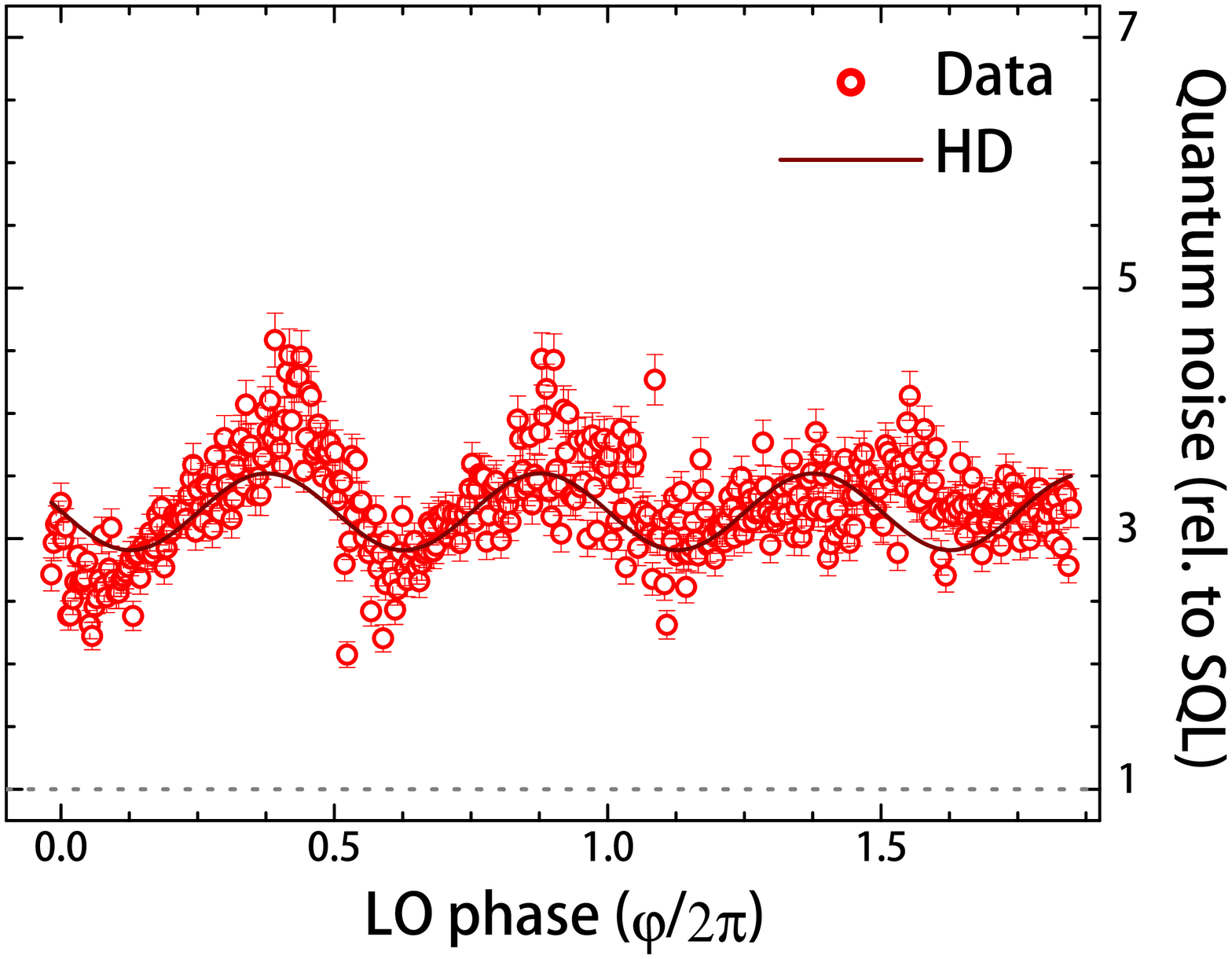}
	\newline
\includegraphics[width=0.6\linewidth]{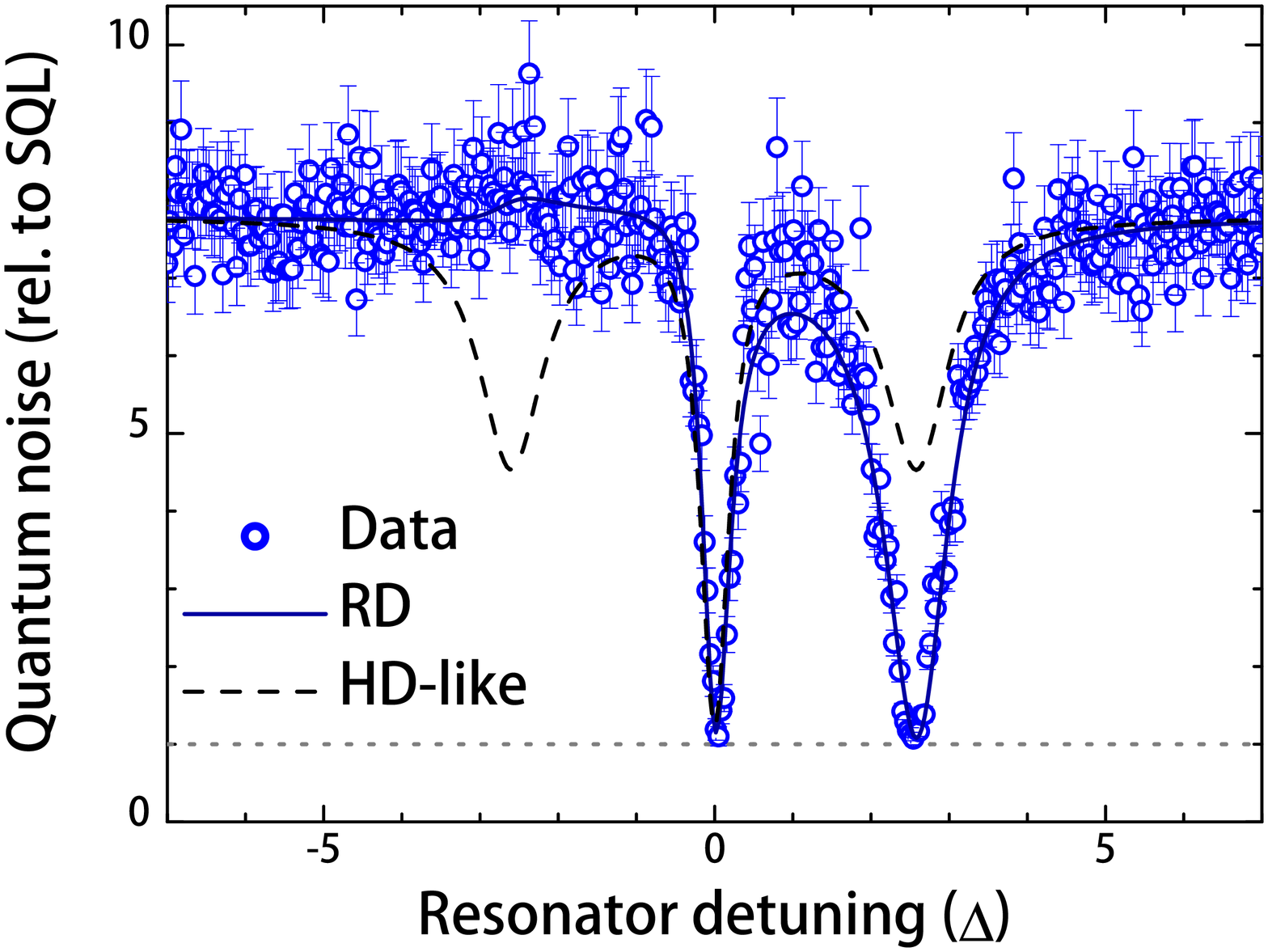}
\hspace{3pt}
\raisebox{1.0cm}{\includegraphics[width=0.35\linewidth]{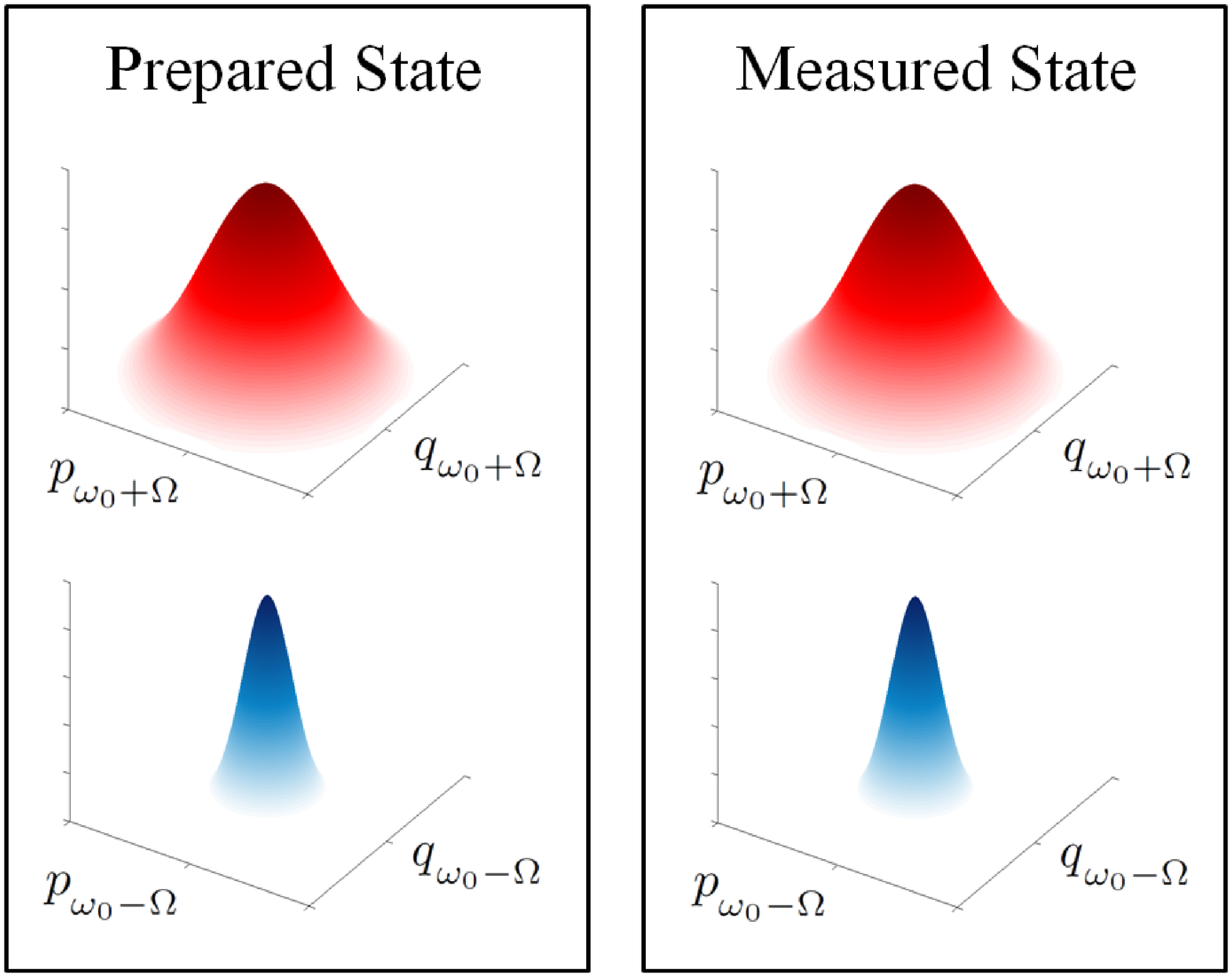}}
\caption{(Color online) Measurements of spectral quantum noise produced by the imbalanced quantum state $\rho$. Curves and labels follow Fig.~\ref{fig:data1Sym}. }
\label{fig:data1Asym}
\end{figure}

Next, measurements of the benchmark quantum state $\hat\rho$ are shown in Fig.~\ref{fig:data1Asym} for both techniques.
Homodyne detection (top right) retrieves the same qualitative shape of quantum noise as in Fig.~\ref{fig:data1Sym}, as emphasized by theoretical curve fittings of Eq.~(\ref{eq:imixhd}) to the data. Essentially, single-mode subspaces of $\hat\rho$ appear to homodyne detection as balanced attenuated versions of $\hat\rho_r$. Insets in the top left show that two different thermal states are mistakenly identified as carrying same mean energy, an assumption tacitly made whenever quantum noise is observed with spectral homodyne detection. The technique is insensitive to energy imbalance between spectral modes. 

Data acquired with resonator detection appear in the bottom left. The solid line represents a theoretical fitting of Eq.~(\ref{eq:SRDenergy}) to the data. The dashed line shows quantum noise as it would appear if only the moments measured with homodyne detection were available, for comparison. Equivalently, this also shows how a quantum state showing balanced modal energy would produce a symmetric noise curve around resonance. The solid and dashed curves show a strong disagreement, establishing the need to invoke strong modal energy imbalance to explain the data on $\hat\rho$. The experimental signature is clear at $\Delta \approx\pm2.6$, when the optical resonator replaces one of the spectral modes of interest by vacuum. At $\Delta \approx -2.6$, the resonator reflects the populated spectral mode and transmits the attenuated mode, causing no perceptible change to quantum noise. The situation is reversed, however, at $\Delta \approx 2.6$, and a large dip can be seen in quantum noise. The Wigner functions show that both $\hat\rho_r$ and $\hat\rho$ are correctly identified with resonator detection. 

In conclusion, we have shown that resonator detection allows one to collect more information about the two-mode spectral quantum state than homodyne detection. In particular, resonator detection is able to detect energy imbalances between sideband modes' fluctuations. Any quantum state spanning over the two-sideband modes can be exactly reconstructed. As a complete measurement technique, it also allows an experimental determination of state purity, in contrast to the intrinsically imperfect spectral homodyne detection. In order to efficiently implement quantum information protocols with continuous variables, one needs to use superior measurement techniques, such as resonator detection. 

\acknowledgments
We thank Y. Golubev and T. Golubeva for stimulating discussions in the initial stage of this research. 
 This work was supported by grants \#2010/52282-1, \#2010/08448-2, \#2009/52157-5, S\~ao Paulo Research Foundation (FAPESP), CNPq, CAPES (through program PROCAD), and CNRS. CF is a member of the Institut Universitaire de France. This research was performed within the framework of the Brazilian National Institute for Science and Technology in Quantum Information (INCT- IQ).

\end{document}